\documentclass[aps,prd,twocolumn,groupedaddress,nofootinbib,longbibliography]{revtex4-1}

\usepackage{amssymb}
\usepackage{amsmath}
\usepackage[linktocpage=true]{hyperref}
\usepackage{graphicx}

\begin{document}

\title{Schwarzschild black hole encircled by a rotating thin disc:\\
       Properties of perturbative solution}

\author{P. Kotla\v{r}\'{\i}k}
\email[]{kotlarik.petr@gmail.com}

\author{O. Semer\'ak}
\email[]{oldrich.semerak@mff.cuni.cz}

\author{P. \v{C}\'{\i}\v{z}ek}
\email[]{ciz@matfyz.cz}

\affiliation{Institute of Theoretical Physics, Faculty of Mathematics and Physics, Charles University, Prague, Czech Republic}

\date{\today}

\begin{abstract}
Will \cite{Will-74} solved the perturbation of a Schwarzschild black hole due to a slowly rotating light concentric thin ring, using Green's functions expressed as infinite-sum expansions in multipoles and in the small mass and rotational parameters. In a previous paper \cite{CizekS-17}, we expressed the Green functions in closed form containing elliptic integrals, leaving just summation over the mass expansion. Such a form is more practical for numerical evaluation, but mainly for generalizing the problem to extended sources where the Green functions have to be integrated over the source. We exemplified the method by computing explicitly the first-order perturbation due to a slowly rotating thin disc lying between two finite radii. After finding basic parameters of the system -- mass and angular momentum of the black hole and of the disc -- we now add further properties, namely those which reveal how the disc gravity influences geometry of the black-hole horizon and those of circular equatorial geodesics (specifically, radii of the photon, marginally bound and marginally stable orbits). We also realize that, in the linear order, no ergosphere occurs and the central singularity remains point-like, and check the implications of natural physical requirements (energy conditions and subluminal restriction on orbital speed) for the single-stream as well as counter-rotating double-stream interpretations of the disc.
\end{abstract}


\maketitle

\section{Introduction}

Disc accretion onto a black hole is a likely engine of some of the most energetic astrophysical sources, like active galactic nuclei, X-ray binaries or gamma-ray bursts. Modelling of such sources naturally starts from determination of the gravitational field of the black hole encircled by a disc. Although the disc typically has just a tiny fraction of the black-hole mass, its gravitational effect may be important on the level of higher derivatives of the potential (space-time curvature), which in turn are crucial for stability of motion of its own matter. To take the disc's gravity into account is however only easy in a static (non-rotating) and axially symmetric case when exact ``superposition'' of the two sources can at least partially be obtained analytically. The simple static setting has been studied many times in the literature; it {\em can} approximate some properties of the actual accretion systems, but it lacks a significant feature suggested by every depiction of accretion onto compact objects: rotation.

If the rotation of the central black hole is slow, one of the analytical options is to perform a small perturbation of a Schwarzschild solution and adjust it to the boundary conditions corresponding to a chosen source -- usually a ring, a disc or a toroid. This means to expand the relevant quantities in the Einstein equations in corresponding small parameters (typically related to mass and angular momentum of the additional matter) and then try to solve the equations to a desired order of expansion. In the previous paper \cite{CizekS-17} we considered a linear (first-order) perturbation of the Schwarzschild black hole due to a slowly rotating concentric finite thin disc. Inspired by \cite{Will-74} who calculated a perturbation due to an infinitesimally thin ring, we expressed in closed form the Green functions (for metric functions representing gravitational potential and rotational dragging), given in his paper as series in orthogonal polynomials. The closed form uses elliptic integrals and is more practical for numerical evaluation, but mainly for studying extended sources when the Green functions have to be integrated over the source volume. We illustrated the method on linear perturbation due to a simple disc existing between two finite radii and having constant Newtonian density.

In the cited paper we provided a longer introduction and described there thoroughly the perturbation method as well as interpretation of the disc both in terms of a one-component ideal fluid and in terms of a two-component (counter-rotating) geodesic dust. Let us thus only remind that the general metric considered is that of circular, i.e. stationary and axisymmetric plus orthogonally transitive space-times,
\begin{align}
  {\rm d}s^2 = &-e^{2\nu}{\rm d}t^2+B^2 r^2 e^{-2\nu}\sin^2\theta\;({\rm d}\phi-\omega{\rm d}t)^2+ \nonumber \\
               &+e^{2\zeta-2\nu}({\rm d}r^2+r^2{\rm d}\theta^2) \,,
  \label{metric}
\end{align}
where the unknown functions $\nu$, $B$, $\omega$ and $\zeta$ depend only on $r$ and $\theta$ covering the meridional surfaces. Note that the above coordinates (mostly called isotropic) are related to the Weyl-type cylindrical coordinates $\rho$ and $z$ by
\[\rho=r\sin\theta, \qquad z=r\cos\theta.\]
The unknown functions are given by Einstein equations with appropriate boundary conditions.
The function $B$ can in our case be chosen to read
\begin{equation}  \label{B-choice}
  B = 1-\frac{k^2}{4r^2}
    = \frac{(2r-k)(2r+k)}{4r^2}\;,
\end{equation}
which corresponds to a horizon lying on $r\!=\!k/2$ (namely where $B\!=\!0$); this also provides the meaning of the parameter $k$. We start from a Schwarzschild background,
\begin{align*}
  {\rm d}s^2= & -\left(\frac{2r-M}{2r+M}\right)^{\!2}{\rm d}t^2+ \\
              & +\frac{(2r+M)^4}{16r^4}
                 \left[{\rm d}r^2+r^2({\rm d}\theta^2+\sin^2\theta\,{\rm d}\phi^2)\right],
\end{align*}
so in our case $k$ coincides with the black-hole mass $M$.
Note that the Schwarzschild background is described by
\begin{equation}
  \nu_0\equiv\nu_{\rm Schw} = \ln\frac{2r-M}{2r+M}
\end{equation}
and that the radial derivative of this potential (often occurring in calculations) reads, with our choice (\ref{B-choice}) for $B$,
\[\nu_{0,r}=\frac{MB}{r^2} \,.\]

\begin{figure*}[ht]
\vspace*{14mm}
\begin{center}
\includegraphics[width=0.74\textwidth]{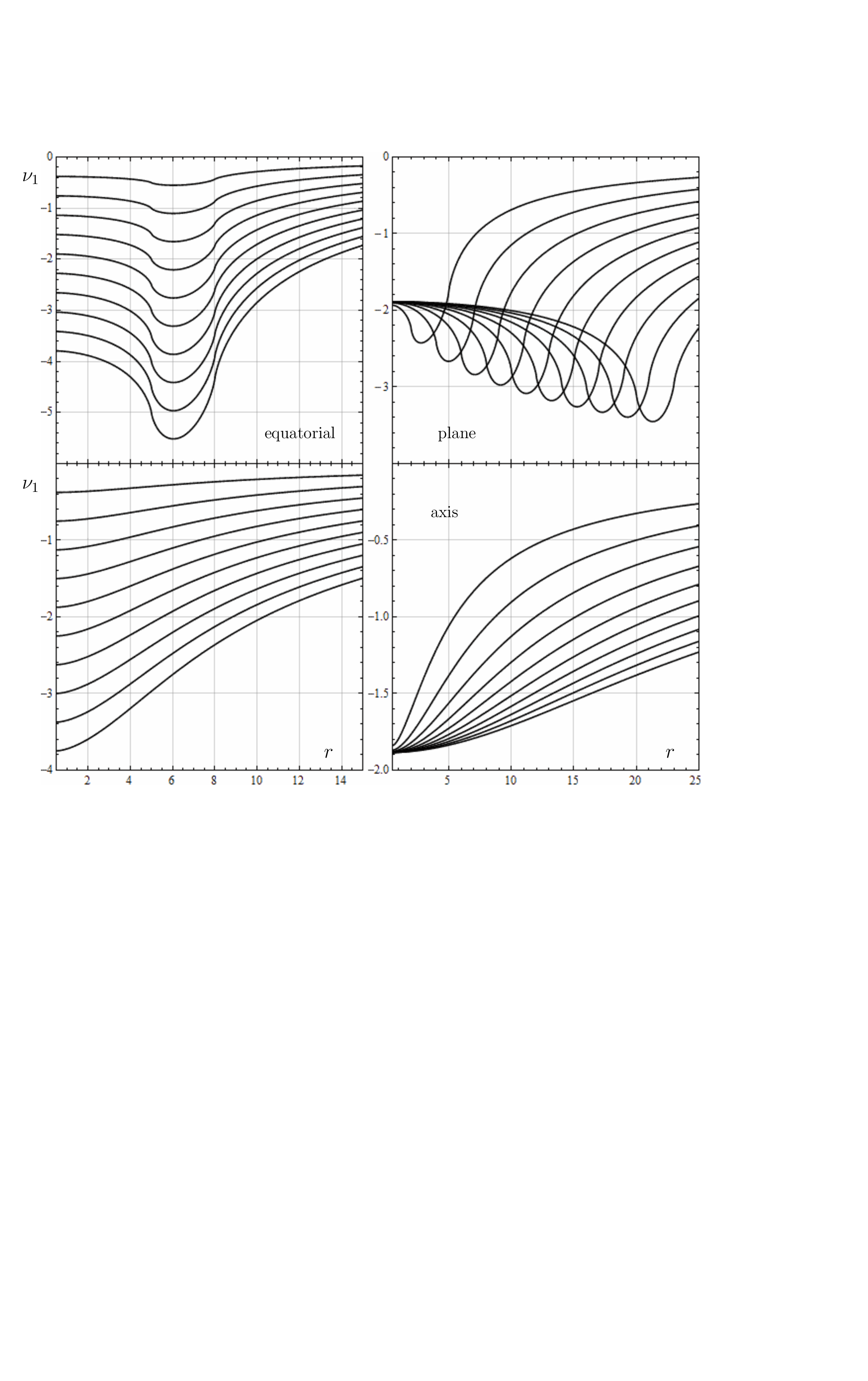}
\caption
{Radial profiles of the constant-density-disc potential, plotted in the disc ($\equiv$ equatorial) plane (top row) and along the symmetry axis (bottom row). In the left column, the disc stretches between the radii $r_{\rm in}\!=\!5M$ and $r_{\rm in}\!=\!8M$ and its density is (from top to bottom curve) $S\!=\!0.02/M$, $0.04/M$, $0.06/M$, \dots, $0.2/M$. In the right column, the density is set to $S\!=\!0.1/M$ and the disc has radial width $3M$, with inner radius (from top to bottom curve again) $r_{\rm in}\!=\!2M$, $4M$, $6M$, \dots, $20M$. The potential clearly behaves according to expectation. (There is no black hole included in these plots, so the parameter $M$ can be understood just as a certain mass scale.)}
\end{center}
\label{nu1-equat-axis}
\end{figure*}

The main task of any stationary and axisymmetric problem is to find the functions $\nu$ and $\omega$ which have the meaning of a gravitational potential and of a dragging angular velocity, respectively. We derived the perturbative (linear-order) solution for these functions in the first paper \cite{CizekS-17}, in particular, we expressed in closed form the Green functions for both the functions and illustrated, on a simple example of a disc with constant Newtonian density stretching between two finite radii $r_{\rm out}\!>\!r_{\rm in}\!>\!M/2$, how the Green functions can be employed to obtain a solution for an extended source. Because the solution is quite cumbersome, we will not repeat the formulas here, just asking the reader to see equations (83) and (81) in the above paper for the perturbation of $\nu$ (denoted by $\nu_1$ and representing entirely the effect of the disc) and equations (84)--(86) for the perturbation of $\omega$ (denoted by $\omega_1$ and equal to the total $\omega$). The solution depends, besides the black-hole mass $M$, on four parameters: the inner and outer radius of the disc ($r_{\rm in}$ and $r_{\rm out}$ in terms of the isotropic radius) and two densities, one (denoted by $S$) having the meaning of Newtonian surface mass density and scaling the disc mass and (thus) potential, and the other (denoted by $W$) having similar role for the dragging function $\omega$. Examples of the disc-potential profiles are shown in Figure \ref{nu1-equat-axis}, as plotted in the equatorial plane (given by the disc) and along the symmetry axis.

In the previous paper we found parameters of the one-component as well as two-component interpretation of the resulting disc (density and pressure in the former case, while two densities in the latter one, plus the corresponding velocities). We also computed the mass and angular momentum of the black hole and of the disc (see equations (113) and (114) of Paper I), the main point being that we adjusted the black-hole rotation (which can be chosen rather freely within the linear order) in such a manner that the hole keeps, in perturbation, its mass as well as zero angular momentum, while the angular velocity of its horizon becomes non-zero (positive). This means that the hole is just being ``dragged along'' by the disc.\footnote
{The freedom in the choice of the black-hole angular momentum arises because dragging ($\omega$) enters the potential ($\nu$) only in the second order, so in the first order it has no back effect on the potential.}
Asking about total mass and angular momentum, we found the asymptotic behaviour of $\nu$ and $\omega$ at radial infinity. We also checked how the solution looks at important locations, namely on the axis, in the equatorial plane and on the horizon.

The present paper is devoted to further properties of the perturbative black-hole--disc solution. Section \ref{horizon} deals with the influence of the disc on geometry of the black-hole horizon, as revealed by isometric embedding of its meridional outline into $\mathbb{E}^3$, by the latter's Gauss curvature, by its proper area and surface gravity, and by its equatorial and meridional proper circumferences. Then, in Section \ref{static-limit,singularity}, we briefly realize that the first-order perturbation does not give rise to an ergosphere and that the central singularity remains given by that of the original, Schwarzschild space-time. Section \ref{circular} focuses on the influence of the disc on the properties of circular equatorial geodesics, in particular on the positions of the photon, marginally bound and marginally stable orbits, as well as on that of a ``marginally possible'' (or ``Lagrangian'') orbit where the gravitational influence from the hole and from the external source just compensate. Finally, in Section \ref{phys-conditions}, we check implications of several natural physical requirements (energy conditions and subluminal character of the disc-matter motion) for interpretations of the disc in terms of a one-component fluid as well as in terms of a two-component dust.

We use geometrized units in which $c\!=\!1$, $G\!=\!1$, metric tensor $g_{\mu\nu}$ has the signature ($-$$+$$+$$+$), Greek indices run from 0 to 3 and index-posed comma denotes partial derivative.

\subsection{Validity of the linear perturbation order}
\label{validity}

Restricting to the linear perturbation order means that during derivation of the solution one neglects all terms quadratic and higher-order in the perturbation quantities $\nu_1$ and $\omega_1\!\equiv\!\omega$ as well as in any of its derivatives. Validity of such a result is, roughly speaking, restricted to regions where $\nu_1$ and $\omega$ as well as their derivatives are small with respect to the unperturbed potential $\nu_0$. Practically this depends on where the disc is placed -- if its radius is large, it lies where the black-hole influence is already weak, so even a very low values of the densities $S$ and/or $W$ can make its effect dominant, mainly in its vicinity. However, one is clearly more interested in the case when the disc lies on radii where astrophysical accretion discs are supposed to have their inner radii, i.e. around or somewhat below $10M$. It is of course simple to evaluate and compare $\nu_1$, $\omega$ (and their derivatives) with $\nu_0$ for any given set of parameters, but let us only restrict here to providing {\em some} idea by saying that for a disc lying -- in terms of the isotropic radius $r$ -- between $r_{\rm in}\!=\!5M$ and $r_{\rm out}\!=\!8M$ (which is just above the pure-Schwarzschild radius of the innermost stable circular orbit), the linear approximation is valid up to some $S\!\simeq\!0.002/M$ and up to some $W\!\simeq\!20M$: for such values, the disc potential $|\nu_1|$ and the dragging function $|\omega|$ are at worst (close to the disc) about $3\times$ smaller than the black-hole potential $|\nu_0|$, so the neglected quadratic terms are about $10\times$ smaller. Let us add that $S\!\simeq\!0.002/M$ and $W\!=\!20/M$ imply, for the $5M\!<\!r\!<\!8M$ disc, that its mass is about ${\cal M}_1\!\doteq\!0.25M$ and its angular momentum is about ${\cal J}_1=7.45M^2$. We checked that a similar bound also holds for gradient of $\omega$ (this is important, because it is the gradient squared through which $\omega$ enters the equation for $\nu$ -- see equation (6) or (12) in Paper I); however, dragging generally falls off much faster than potential when receding from the source, and one has to be careful if using higher (than the first) derivatives of $\omega$ -- these are typically larger, at least close to the disc (especially close to its edges).

A minor note concerning the term $e^{\nu_1}$ which frequently appears in the formulas below: the perturbation of the potential $\nu_1$ is one of the quantities which should be left only in linear order, yet we often do {\em not} expand $e^{\nu_1}$ to $(1\!+\nu_1)$ (though it can of course be done easily). Namely, since the dragging ($\omega$) only ``back-affects'' the potential in the second perturbation order, the properties which do not contain $\omega$ (in the linear order) behave like in the {\em exact} static axisymmetric case. If -- for example for the illustrations (of such properties) to better show some tendency -- one also admits larger values of $\nu_1$ than those strictly complying with the linear regime, it is thus more safe to use the full formula than its linear-in-$\nu_1$ part. (The result is anyway relevant only where the linear approximation is acceptable.)

\section{Geometry of the horizon}
\label{horizon}

With the choice $B\!=\!1\!-\!\frac{k^2}{4r^2}\,$, the black-hole horizon lies at $r\!=\!k/2$ (where $k\!=\!M$ in our case, since we start from the Schwarzschild metric), so its coordinate picture is exactly spherical irrespectively of the perturbation. However, the intrinsic shape of the horizon (given by proper distances in the two angular directions) does change due to the presence of the additional source. In order to reveal this, let us take the horizon as the two-dimensional surface $\{t\!=\!{\rm const},r\!=\!M/2\}$. From (\ref{metric}), its metric reads
\begin{align*}
  {\rm d}s^2_{\rm H}
  &= (g_{\theta\theta})_{\rm H}\,{\rm d}\theta^2 + (g_{\phi\phi})_{\rm H}\,{\rm d}\phi^2 = \\
  &= \frac{M^2}{4}\,(e^{2\zeta-2\nu})_{\rm H}\,{\rm d}\theta^2
    +\frac{M^2}{4}\,(B^2 e^{-2\nu})_{\rm H}\sin^2\theta\,{\rm d}\phi^2 = \\
  &= \frac{M^2}{4}\left[\frac{B^2}{e^{2\nu(0)}}\right]_{\rm \!H}
     \left[\frac{e^{2\nu(\theta)}}{e^{2\nu(0)}}\,{\rm d}\theta^2
           +\frac{e^{2\nu(0)}}{e^{2\nu(\theta)}}\,\sin^2\theta\,{\rm d}\phi^2\right]_{\rm \!H},
\end{align*}
where the relation valid on stationary and axisymmetric horizon
\begin{equation}  \label{zetaH}
  \zeta_{\rm H}(\theta)=2\nu_{\rm H}(\theta)-2\nu_{\rm H}(0)+\ln B
\end{equation}
has been employed (see e.g. \cite{Will-74}, equation (24)).
Substituting now $\nu=\nu_0+\nu_1$, where $\nu_0\!\equiv\!\nu_{\rm Schw}$ and $\nu_1$ is the perturbation brought by the external source, one has
\[(B^2 e^{-2\nu})_{\rm H}=16 e^{-2\nu_1}\]
and the Schwarzschild part of the exponent $4\nu_{\rm H}(\theta)\!-\!4\nu_{\rm H}(0)$ cancels out,
so the horizon metric reduces to
\begin{equation}  \label{metric,H}
  {\rm d}s^2_{\rm H}
  = \frac{4M^2}{e^{2\nu_1(0)}}
    \left[\frac{e^{2\nu_1(\theta)}}{e^{2\nu_1(0)}}\,{\rm d}\theta^2
           +\frac{e^{2\nu_1(0)}}{e^{2\nu_1(\theta)}}\,\sin^2\theta\,{\rm d}\phi^2\right]
\end{equation}
evaluated at $r\!=\!M/2$.
Hence, the horizon geometry depends only on the metric function $\nu$ and thus in the first perturbation order it behaves like in the exact static-case superposition. The latter has been solved at many places, see e.g. \cite{Erratum-01}.

\subsection{Isometric embedding into $\mathbb{E}^3$}

In that paper (actually an erratum), the isometric embedding was summarized (as taken from \cite{Smarr-73}) of the horizon two-surface in a three-dimensional Euclidean space (revealing the actual horizon shape), and also the prescription for its Gauss curvature was given. Let us just briefly recall that the isometric imbedding starts from writing the metric in the form
\[{\rm d}s^2 = \eta^2\left[f^{-1}(\mu)\,{\rm d}\mu^2+f(\mu)\,{\rm d}\phi^2\right],\]
where $\mu\!:=\!\cos\theta$ and, in our case,
\begin{align*}
  \eta   &= 2M e^{-\nu_1(\mu=1)}, \\
  f(\mu) &= (1\!-\!\mu^2)\,e^{2\nu_1(\mu=1)-2\nu_1(\mu)},
\end{align*}
with $\nu_1$ already understood to be evaluated at the horizon ($r\!=\!M/2$).
The embedding into a Euclidean three-space endowed with Cartesian coordinates $(x,y,z)$ is then given by
\begin{equation}
  \frac{x}{\eta}=\sqrt{f}\,\cos\phi, \;\;
  \frac{y}{\eta}=\sqrt{f}\,\sin\phi, \;\;
  \frac{z}{\eta}=\int\limits_0^\mu\sqrt{\frac{4-(f_{,\mu})^2}{4f}}\,{\rm d}\mu \,.
\end{equation}

\begin{figure}[ht]
\vspace*{9mm}
\includegraphics[width=\columnwidth]{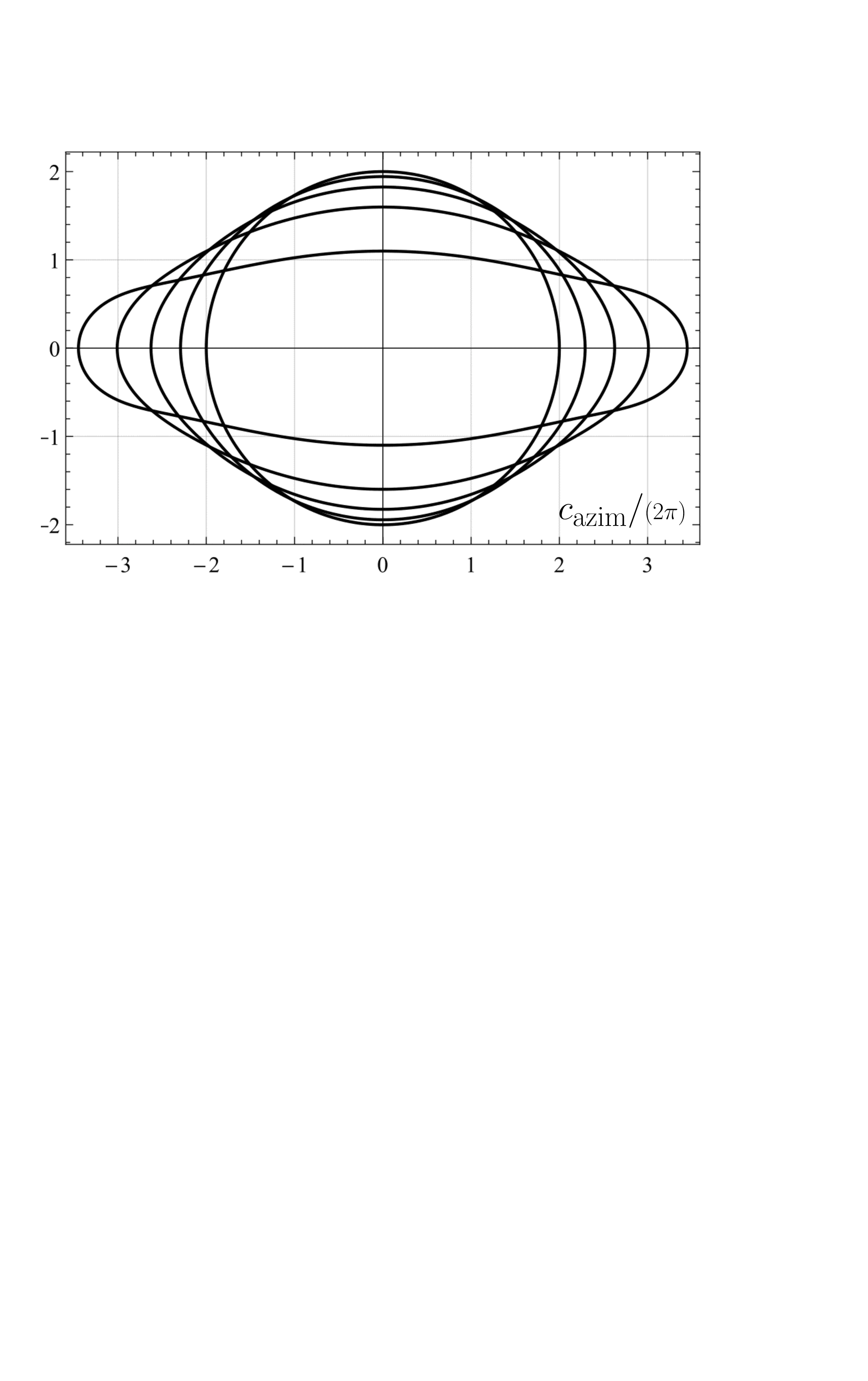}
\caption
{Meridional section ($\phi\!=\!{\rm const}$) of the horizon's isometric embedding into $\mathbb{E}^3$, with the symmetry axis going in the vertical direction and the equatorial plane perpendicular to it. The horizontal axis thus represents azimuthal circumferential radius (proper azimuthal circumference of the horizon divided by $2\pi$), while in the vertical direction the contour goes in such a way that its length represents proper distance measured along the horizon in the meridional (polar) direction. For a disc lying between $r_{\rm in}\!=0.8333M$ and $r_{\rm out}\!=\!1M$ and having density $S\!=\!0.0$, $0.1$, $0.2$, $0.3$ and $0.4$ (in the units of $1/M$), the horizon becomes more and more flattened. Both axes are in the units of $M$.}
\label{horizon-shape}
\end{figure}

\subsection{Gauss curvature}

The Gauss curvature is itself a good indicator of how the horizon behaves when subjected to a tidal effect of the external source. In particular, it is a common experience that, when the source is about the equatorial plane, the horizon's axial parts may be so ``strained'' that the Gauss curvature decreases below zero there (a similar distortion also arises as a consequence of rotation, as in the Kerr case). The Gauss curvature equals half of the Ricci curvature scalar, computed for the chosen two-dimensional surface. For the metric (\ref{metric}), the $r\!=\!{\rm const}$ surface has Gauss curvature
\[K(r\!=\!{\rm const})
  =\frac{1+\nu_{,\theta\theta}+(\nu_{,\theta}+\zeta_{,\theta})\cot\theta-\nu_{,\theta}\zeta_{,\theta}}
        {r^2\,e^{2\zeta-2\nu}} \;,\]
which specifically for the horizon ($r\!=\!M/2$, $\zeta_{,\theta}\!=\!2\nu_{,\theta}$) gives
\begin{equation}
  K_{\rm H}=\frac{1+\nu_{1,\theta\theta}+3\nu_{1,\theta}\cot\theta-2(\nu_{1,\theta})^2}
                 {4M^2\,e^{2\nu_1(\theta)-4\nu_1(0)}} \;,
\end{equation}
where $\nu_1$ again represents just the external source. (The last, quadratic term should therefore be omitted in order to comply with the first-order approximation, and the exponential downstairs can also be expanded accordingly.)
Like for the isometric embedding, the explicit expression valid for our constant-density disc is rather cumbersome, but on the axis it reduces to
\begin{equation}  \label{Gauss,axis}
  K_{\rm H}(\theta\!=\!0) = \frac{1-8\pi MS\left(\frac{1}{x_{\rm in}}-\frac{1}{x_{\rm out}}\right)}
                                 {4M^2\,e^{4\pi MS(x_{\rm out}-x_{\rm in})}}  \;,
\end{equation}
where we have used the horizon value (equation (90) in the previous paper)
\begin{equation}  \label{nu1,x=1}
  \nu_1(x\!=\!1)=
  -2\pi MS\left(\!\sqrt{x_{\rm out}^2\!-\!\sin^2\theta}-\sqrt{x_{\rm in}^2\!-\!\sin^2\theta}\right),
\end{equation}
as written in the radial variable
\[x:=\frac{r}{M}+\frac{M}{4r}\]
(in terms of the latter, the horizon is on $x\!=\!1$).\footnote
{For our choice $B\!=\!1-M^2/(4r^2)$, it holds $x_{,r}\!=\!B/M$.
It is also useful to note that $Be^{-\nu_0}=(2r+M)^2/(4r^2)$.}
Clearly (\ref{Gauss,axis}) can become negative for sufficiently large Newtonian density of the disc $S$ (and for the outer radius of the disc $x_{\rm out}$ sufficiently larger than the inner radius $x_{\rm in}$). For $S\!=\!0$, the expression reduces to the Schwarzschild value $1/(4M^2)$.

The effect of the constant-density disc on the horizon's intrinsic geometry is illustrated in Figure \ref{horizon-shape}. As clear from above, the effect is solely determined by the external-source potential (the parameter $W$ is thus irrelevant in the linear perturbation order), so in the case of our disc equations (83) and (81) of \cite{CizekS-17} are important. The plot may look almost like a repetition of the figure presented in \cite{Erratum-01}, but it is not so -- the sources considered there were different from the present disc (there, it was a Bach-Weyl thin ring and an infinite disc obtained by inversion of the first member of the Morgan-Morgan counter-rotating family). Anyway, the plot does not need much commenting -- the horizon inflates towards to external source as expected. What could however be mentioned here is the paper by \cite{Lanza-92} who solved the black-hole--thin-disc problem numerically and obtained, in some {\em slowly-rotating} cases, a {\em prolate} horizon (see discussion at the very end of our previous paper \cite{CizekS-17}). Such an observation has not been repeated in any other study, and there also does not seem to be any chance for it in our case. Actually, this is already clear from equation (\ref{metric,H}): the prolate-horizon eventuality would require $\nu_1(\theta)\!>\!\nu_1(0)$ (the potential well generated by the external-source would have to be deeper at the horizon's poles than elsewhere), which, for an attractive equatorial source, is never true.

\begin{figure}[ht]
\vspace*{9mm}
\includegraphics[width=\columnwidth]{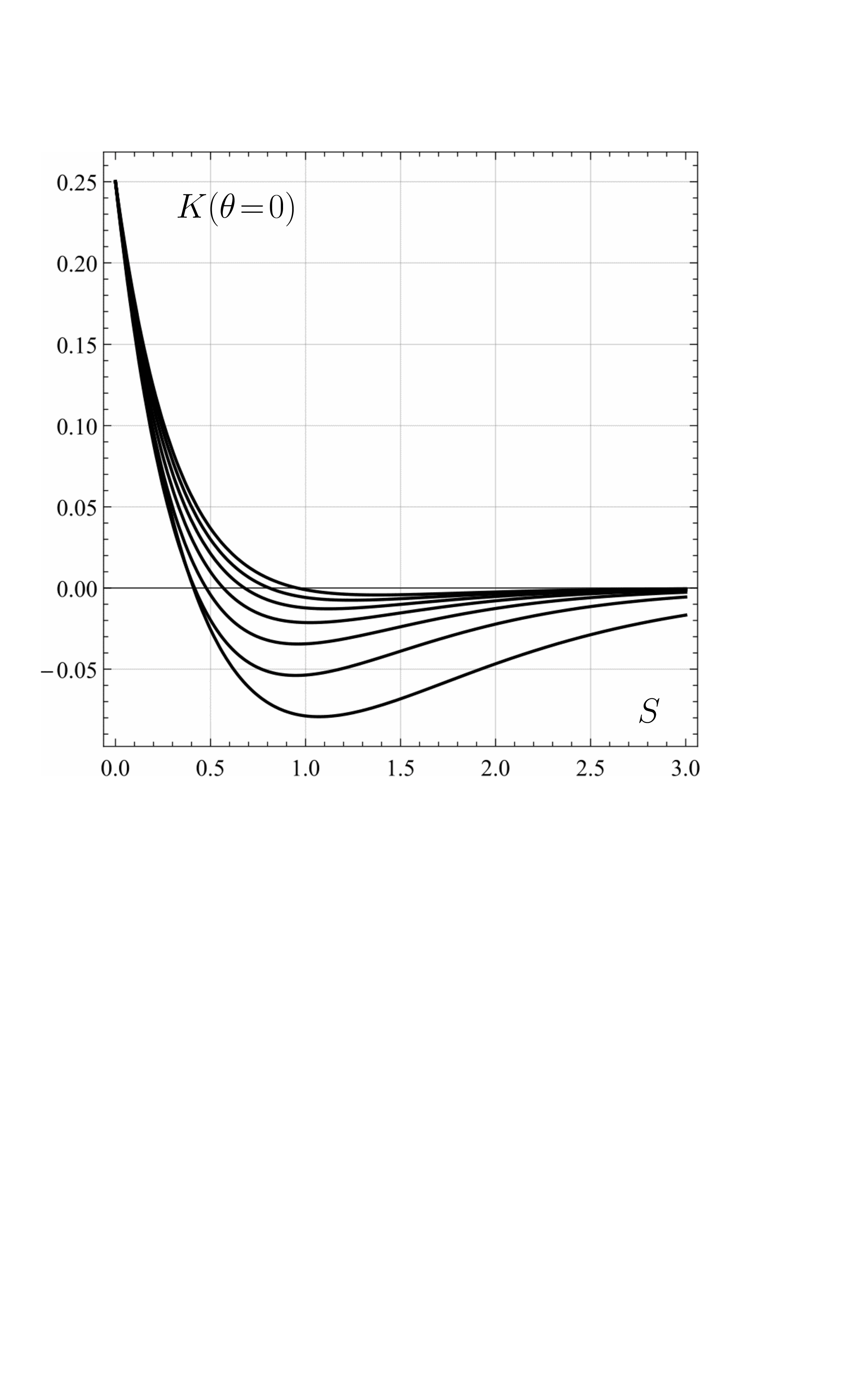}
\caption
{Gauss curvature of the horizon at the axis of symmetry ($\theta\!=\!0$) in dependence on the disc density $S$, plotted for discs of radial width $r_{\rm out}\!-r_{\rm in}\!=\!0.2M$ and having -- from the bottom to the top curve -- inner radius at $r_{\rm in}\!=\!0.7M$, $0.9M$, $1.1M$, \dots, $1.9M$. For all of these, the Gauss curvature falls to negative values if the disc is sufficiently dense/massive. Axes are in the units of $1/M$ and $1/M^2$, respectively.}
\label{Gauss-axis}
\end{figure}

Figure \ref{Gauss-axis} shows how the Gauss curvature of the horizon at the axis pole behaves in dependence on the disc density $S$. Several cases with different inner disc radii (while the same width $r_{\rm out}\!-r_{\rm in}$) are plotted. Generally, the curvature decreases from the Schwarzschild value of $1/(4M^2)$ and eventually falls below zero when the disc density (and thus mass) is increased. Again, the plot resembles figure 2 of \cite{Semerak-03} where the same effect was studied for an exact superposition of a Schwarzschild black hole with various members of the inverted Morgan-Morgan counter-rotating disc family.

\subsection{Proper area and surface gravity}

Another horizon properties on which the external-source effect can be studied are the horizon's proper area and surface gravity. The area is given by 
\[A_{\rm H} = \oint_{\rm H}\sqrt{g_{\theta\theta}g_{\phi\phi}}\;{\rm d}\theta\,{\rm d}\phi
            = 2\pi\int_0^\pi (Br^2 e^{\zeta-2\nu})_{\rm H}\,\sin\theta\,{\rm d}\theta \,\]
and in our case -- with (\ref{zetaH}) and (\ref{nu1,x=1}) -- amounts to
\begin{equation}
  A_{\rm H}=16\pi M^2 e^{-2\nu_1(x=1,\theta=0)}
           =16\pi M^2 e^{4\pi MS(x_{\rm out}-x_{\rm in})}.
\end{equation}

The surface gravity on a stationary and axisymmetric horizon is given by
\[\kappa_{\rm H}^2:=\lim\limits_{N\rightarrow 0^+}(g^{\mu\nu}N_{,\mu}N_{,\nu})
                   =\left\{e^{4\nu-2\zeta}\left[(\nu_{,r})^2\!+\!\frac{(\nu_{,\theta})^2}{r^2}\right]\right\}_{\rm \!H}\!,\]
where the lapse function $N$ is simply $N\!\equiv\!e^\nu$. Substituting again our case, we have
\begin{equation}
  \kappa_{\rm H}=\frac{e^{2\nu_1(x=1,\theta=0)}}{4M}
                =\frac{1}{4Me^{4\pi MS(x_{\rm out}-x_{\rm in})}}
                =\frac{4\pi M}{A_{\rm H}} \,.
\end{equation}
Note that this result is independent of $\theta$ ($\kappa_{\rm H}$ is uniform all over the horizon) as it should be, according to the zeroth law of black-hole thermodynamics, for any stationary horizon.
Clearly the horizon area grows rapidly (exponentially) while the surface gravity falls down when the disc density $S$ increases.
See Figure \ref{AH,kappaH,circum} for illustration.

\subsection{Equatorial and meridional circumferences}

Deformation of the horizon due to the external source is naturally accompanied by a change of the ratio between its equatorial and meridional proper circumferences. The equatorial (actually any azimuthal) one is very simple due to the axial symmetry,
\begin{align*}
  c_{\rm equa}&=2\pi\sqrt{g_{\phi\phi}(r\!=\!M/2,\theta\!=\!\pi/2)}= \\
              &=\pi M\,\left[Be^{-\nu(\theta=\pi/2)}\right]_{\rm H},
\end{align*}
and in our black-hole--disc case it comes out
\begin{align}
  c_{\rm equa}&=4\pi Me^{-\nu_1(x=1,\theta=\pi/2)}= \nonumber \\
              &=4\pi M\exp\!\left[2\pi MS\left(\!\sqrt{x_{\rm out}^2\!-\!1}-\sqrt{x_{\rm in}^2\!-\!1}\right)\right].
\end{align}
The meridional (in the given coordinates, it means latitudinal) circumference
\begin{align*}
  c_{\rm meri}&=2\int_0^\pi\sqrt{g_{\theta\theta}(r\!=\!M/2)}\;{\rm d}\theta
               =M\int_0^\pi (e^{\zeta-\nu})_{\rm H}{\rm d}\theta = \\
              &=M\left[Be^{-\nu(0)}\right]_{\rm H}
                \int_0^\pi e^{\nu_{\rm H}(\theta)-\nu_{\rm H}(0)}\,{\rm d}\theta
\end{align*}
is more difficult to compute explicitly. For our specific situation, it reads
\begin{equation}
  c_{\rm meri}=4Me^{-2\nu_1(r=M/2,0)}\int_0^\pi e^{\nu_1(r=M/2,\theta)}\,{\rm d}\theta \,,
\end{equation}
with (\ref{nu1,x=1}) substituted for $\nu_1(r\!=\!M/2,\theta)$ again.

Ratio of the horizon's meridional to equatorial circumferences is plotted (together with the proper area and surface gravity) in Figure \ref{AH,kappaH,circum} for several sequences of black-hole--disc configurations, in order to illustrate its dependence on the disc density (mass) and location. The plots correspond to flattening of the horizon due to the disc.\footnote
{Sometimes the behaviour of the $c_{\rm meri}/c_{\rm equa}$ ratio is suggested as a sufficient indicator of the horizon shape. However, this is not always reliable, namely $c_{\rm meri}$ exceeds $c_{\rm equa}$ when the horizon gets stretched along the axis {\em as well as} if it gets concave in the axial regions (while $c_{\rm equa}$ kept constant).}

\begin{figure*}[ht!]
\begin{center}
\vspace*{14mm}
\includegraphics[width=0.67\textwidth]{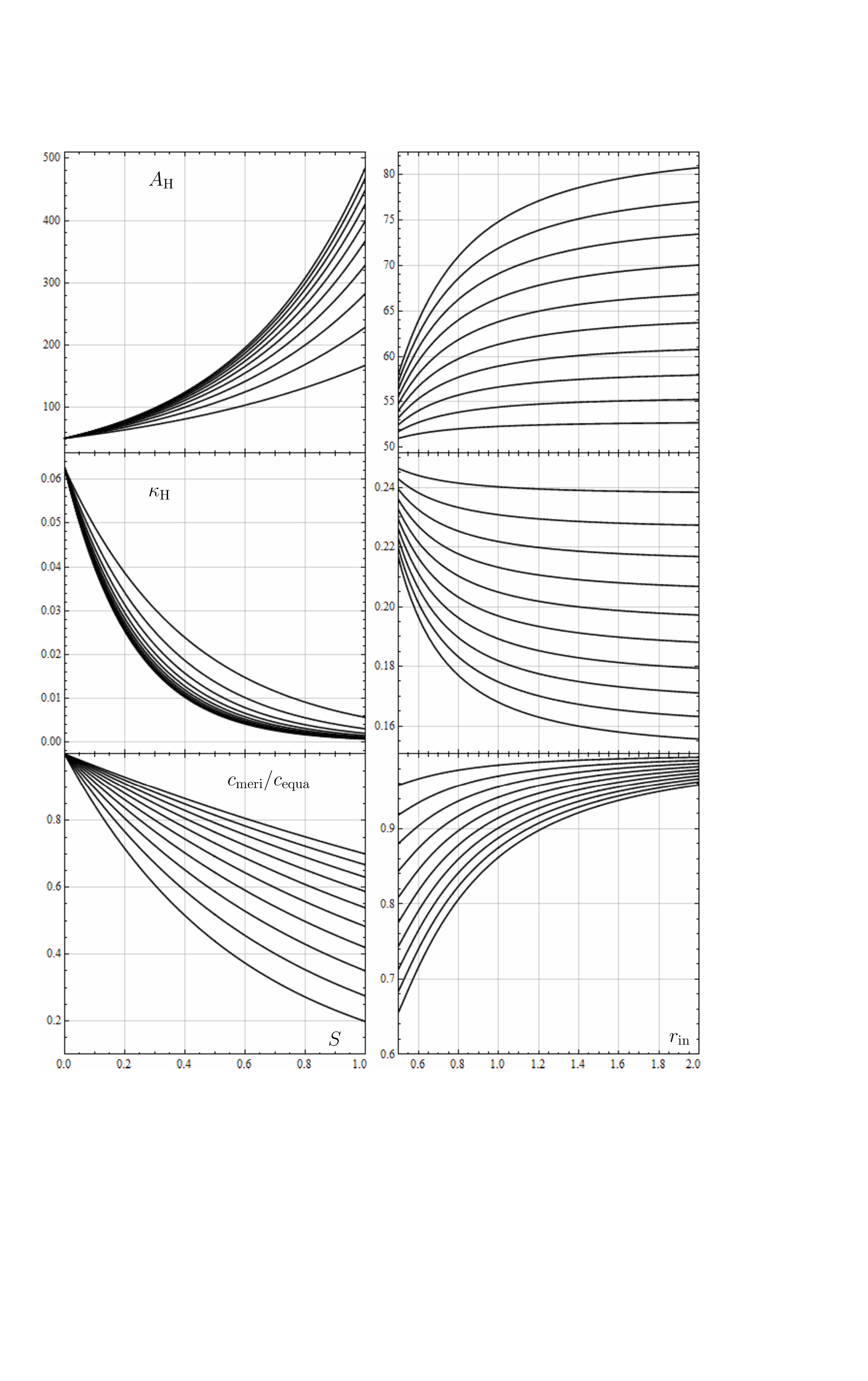}
\caption
{Basic properties of the horizon -- its proper area $A_{\rm H}$ (first row), surface gravity $\kappa_{\rm H}$ (middle row), and ratio of its meridional to equatorial circumferences $c_{\rm meri}/c_{\rm equa}$ (bottom row), plotted for discs of radial width $r_{\rm out}\!-r_{\rm in}\!=\!0.2M$, in dependence on their density $S$ (left column) and inner radius $r_{\rm in}$ (right column). Specifically, in the left plots, the above quantities are plotted, against $S$, for ten different inner radii $r_{\rm in}\!=\!0.6M$, $0.7M$, $0.8M$, \dots, $1.5M$, while in the right plots they are plotted against $r_{\rm in}$ for ten different densities $S\!=\!0.02/M$, $0.04/M$, $0.06/M$, \dots, $0.2/M$. Identification of curves: in the left column, with $S$ growing from zero, $A_{\rm H}$ increases from $16\pi M^2$, $\kappa_{\rm H}$ decreases from $1/(4M)$ and $c_{\rm meri}/c_{\rm equa}$ decreases from 1, the more steeply the smaller is $r_{\rm in}$; in the right plot, with $r_{\rm in}$ growing (from $M/2$, which is the radius of the horizon), $A_{\rm H}$ increases, $\kappa_{\rm H}$ decreases and $c_{\rm meri}/c_{\rm equa}$ increases, the more steeply the larger is $S$.
Units: $[M^2]$ for $A_{\rm H}$, $[1/M]$ for $\kappa_{\rm H}$, dimensionless for $c_{\rm meri}/c_{\rm equa}$, $[1/M]$ for $S$ and $M$ for $r_{\rm in}$.}
\label{AH,kappaH,circum}
\end{center}
\end{figure*}

\section{Static limit and singularity}
\label{static-limit,singularity}

A rotating horizon is usually surrounded by a static limit -- a surface which limits the possibility to stay at rest relative to an asymptotic rest frame (namely to ``resist'' rotational dragging caused by the source). In a non-rotating case, the static limit coincides with the horizon. Our first-order perturbation does {\em not} separate a static limit from the horizon. Actually, if ``staying at rest with respect to infinity'' means to stay in spatial coordinates, i.e. to have four-velocity
\[u^\mu=(u^t,0,0,0) \qquad {\rm with} \qquad u^t=\frac{1}{\sqrt{-g_{tt}}} \;,\]
the static limit is given by $g_{tt}\!=\!0$, which for the metric (\ref{metric}) means
\[-e^{2\nu}+B^2 r^2\omega^2 e^{-2\nu}\sin^2\theta = 0 \,,\]
so in the first order just
\[e^\nu=e^{\nu_0}e^{\nu_1}=0\,.\]
The disc potential is nowhere infinitely deep, $e^{\nu_1}\!>\!0$, so the static limit remains where it lies in the original space-time,
\[e^{\nu_0}=\frac{2r-M}{2r+M}=0
  \qquad \Longleftrightarrow \qquad
  r=\frac{M}{2} \;.\]

Another possible question is whether the physical singularity of the solution is perturbed off its original, point-like character. A detailed answer is difficult and we will not provide it here. First, one should admit that in isotropic coordinates the black-hole interior is not covered. Second, the answer should be reached by identifying possible divergence(s) of (e.g.) the Kretschmann curvature invariant, which for the perturbed metric leads to quite a long expression. Nevertheless, one can judge the answer from the structure of the Kretschmann scalar. Computing this scalar for the metric (\ref{metric}) written in terms of the functions
\[N^2\equiv e^{2\nu}, \; g_{\phi\phi}\equiv B^2 r^2 e^{-2\nu}\sin^2\theta \,, \;
  g_{rr}\equiv e^{2\zeta-2\nu} \;\; {\rm and} \;\; \omega\]
and omitting all the terms quadratic or higher-order in $\omega$, one is left with an expression which does not contain $\omega$ {\em at all}. All the other terms, however, are determined purely by $\nu\!=\!\nu_0\!+\!\nu_1$, of which $\nu_1$ is nowhere singular, so all the singularities of the resulting space-time must be given by singularities of the (Schwarzschild) background.

\section{Properties of equatorial circular geodesics}
\label{circular}

Although there is no self-gravity (non-linear effect of the matter on itself) involved in the first perturbation order, one can still estimate some features on this level already -- simply from how the gravitational field is changed due to the perturbation.
In particular, the modified field implies a modified geodesic structure, ergo a different world-lines of free particles. Indeed, these are being followed by free {\em test} particles, but, in the given approximation, also by the matter which is generating the perturbation. We will focus on important equatorial circular geodesics and check how they are shifted due to the gravity of the additional disc source.

\subsection{Light-like limits of circular motion}

The first important property are the light-like limits of circular motion, namely the light-cone boundaries expressed in terms of the angular velocity $\Omega:={\rm d}\phi/{\rm d}t$, which are given by
\[{\rm d}s^2(\rho\!=\!{\rm const},z\!=\!{\rm const})
  =(g_{tt}+2g_{t\phi}\Omega+g_{\phi\phi}\Omega^2)\,{\rm d}t^2
  =0 \,.\]
Substituting
\[g_{tt}=-e^{2\nu}-g_{\phi\phi}\omega^2, \quad
  g_{\phi\phi}=B^2\rho^2 e^{-2\nu}, \quad
  g_{t\phi}=-g_{\phi\phi}\omega \,,\]
one obtains
\begin{equation}
  \Omega_{\rm min,max}=\omega\mp\frac{e^{2\nu}}{\rho B}
\end{equation}
(where for the equatorial motion everything is to be evaluated at $z\!=\!0$).
In our linear perturbation of Schwarzschild, we have, in the equatorial plane,
\begin{equation}  \label{Omega,min,max}
  \Omega_{\rm min,max}=\omega_1\mp\frac{4r(2r-M)}{(2r+M)^3}\,e^{2\nu_1}.
\end{equation}

\subsection{Zero-speed limit of free circular motion}

The opposite limit of {\em free} circular motion is the case of zero speed. Actually, when speaking of a very compact centre, one immediately imagines high orbital speed, necessary to produce sufficient centrifugal effect to balance the centre's gravity. However, if there is (also) some heavy enough source external to the orbit (the disc in our case), it may attract the test body so strongly that ``no angular velocity is small enough'' (even a body at rest is pulled outwards). Put simply, the test particle has to orbit below the Lagrangian point (in fact a whole circle) of the system.

The limit, Lagrangian orbit, has to lie in the equatorial plane and is given by a very simple condition: the radial acceleration must vanish, i.e., in the linear-perturbation order (when dragging does not enter), $\nu_{,r}(\theta\!=\!\pi/2)\!=\!0$. Substituting $\nu_{\rm Schw}+\nu_1$ for the total potential, one obtains quite a long result (due to $\nu_1$) which is not worth presenting. However, we will evaluate its behaviour numerically and include it in a summarizing figure below.

\subsection{Condition for free circular motion}
\label{free-circular}

The condition that the circular motion be free (geodesic; in astrophysics usually called Keplerian) is easily obtained by demanding that the acceleration corresponding to the four-velocity $u^\mu=u^t(1,0,0,\Omega)$ vanishes. In the Killing-type coordinates, the acceleration has just meridional components, of which the latitudinal (or ``vertical'') one generally vanishes only in the equatorial plane, and vanishing of the radial component has two solutions
\begin{align}
  \Omega_\pm
  &= -\frac{g_{t\phi,\rho}}{g_{\phi\phi,\rho}}\pm
     \sqrt{\left(\frac{g_{t\phi,\rho}}{g_{\phi\phi,\rho}}\right)^{\!2}-\frac{g_{tt,\rho}}{g_{\phi\phi,\rho}}}=
  \nonumber \\
  &= \omega+\frac{g_{\phi\phi}\omega_{,\rho}}{g_{\phi\phi,\rho}}
     \pm\sqrt{\left(\omega+\frac{g_{\phi\phi}\omega_{,\rho}}{g_{\phi\phi,\rho}}\right)^{\!2}
              -\frac{g_{tt,\rho}}{g_{\phi\phi,\rho}}} \;.
  \label{Omega_pm}
\end{align}
In the linear approximation of our problem, clearly the parenthesis under the square root is to be omitted, since it is quadratic in $\omega$. Also, of $g_{tt}\!=\!-e^{2\nu}\!+g_{\phi\phi}\omega^2$ one keeps only the first term, and in the term before the square root one substitutes $\nu\!=\!\nu_0$ into $g_{\phi\phi}\!=\!B^2\rho^2 e^{-2\nu}$. More precisely, the linear approximation here means to take
\begin{align}
  &\frac{g_{\phi\phi}}{g_{\phi\phi,\rho}}
   =\frac{Br^3}{M^2+2Br^2(1-r\,\nu_{0,r})}
   =\frac{r}{2}\,\frac{2r+M}{2r-M}
   =\frac{r}{2e^{\nu_0}} \;, \label{Omega_pm,approx1} \\
  &\sqrt{\left(\omega+\frac{g_{\phi\phi}\omega_{,\rho}}{g_{\phi\phi,\rho}}\right)^{\!2}
              -\frac{g_{tt,\rho}}{g_{\phi\phi,\rho}}}
   \doteq \sqrt{\frac{-g_{tt,\rho}}{g_{\phi\phi,\rho}}}\doteq \nonumber \\
  &\doteq \frac{8\sqrt{Mr^3}\,e^{2\nu_1}}{(2r+M)^3}
          \left(1+\frac{2r+M}{2r-M}\,\frac{4r^2+M^2}{8M}\,\nu_{1,r}\right), \label{Omega_pm,approx2}
\end{align}
where we have already fixed to the equatorial plane (where $\rho\!\equiv\!r$) and indicated by $\doteq$ the restriction to the linear order in $\nu_1$ (or its derivative).
In the Schwarzschild limit, the Keplerian frequencies reduce to
\[\Omega_\pm = \pm\sqrt{\frac{-g_{tt,\rho}}{g_{\phi\phi,\rho}}}
             = \pm\frac{8\sqrt{Mr^3}}{(2r+M)^3} \;.\]

\subsection{Marginally stable circular geodesics}
\label{ms}

For any kind of theoretical behaviour to be realistic, the principal realizability is a necessary condition, but physically not a sufficient one: the behaviour should also be stable. In connection with thin accretion discs, the stability of their orbits with respect to perturbations acting within the disc plane is usually emphasized; it is known to lead to the condition that the angular momentum of circular motion has to increase in the outward radial direction, i.e. that $u_{\phi,r}\!>\!0$. Actually, when subject to such an equatorial perturbation, a circular orbit oscillates, with respect to an asymptotic inertial frame, with the so-called radial epicyclic frequency (see e.g. \cite{SemerakZ-00} for derivation in the same notation) whose square reads
\begin{align}  \label{kappa}
  \kappa^2
  &=\frac{e^{2\nu-2\lambda}}{(u^{t})^{3}u_\phi}\,
    g_{\alpha\phi,\rho}u^{\alpha}g^{t\beta}u_{\beta,\rho}= \nonumber \\
  &=\frac{e^{2\nu-2\lambda}u_{\phi,\rho}}{(u^{t})^{4}\rho^{2}B^{2}}
    \left[u_{\phi,\rho}-(u^{t})^{3}\rho^{2}B^{2}\Omega_{,\rho}\right].
\end{align}
For a {\em geodesic} orbit, the geodesic value(s) of $\Omega$ (\ref{Omega_pm}) should be employed. Let us restrict to the equatorial plane and notice that the result depends on the signs of $u_{\phi,r}$ and $\Omega_{,r}$. Since the perturbation values should be much smaller than the ``background'', Schwarzschild ones, we assume the usual argumentation (valid for isolated stationary black holes) can be applied: $\Omega_+$ decreases and $\Omega_-$ increases with $r$ (namely, their magnitude falls of with distance). Hence, for ``prograde'' orbits, for which $u_\phi\!>\!0$ and $\Omega_{,r}\!<\!0$, stability is ensured by $u_{\phi,r}\!>\!0$ (then $\kappa^2$ comes out positive). For ``retrograde'' orbits, on the other hand, $u_\phi\!<\!0$ and $\Omega_{,r}\!>\!0$, and stability is ensured by $u_{\phi,r}\!<\!0$ (which means that the angular-momentum {\em magnitude} increases).

In order to calculate $u_{\phi,\rho}$, one uses the relations (valid for {\em any} motion in the given type of space-times)
\begin{align*}
  u_\phi &= g_{\phi\phi}u^t(\Omega-\omega) \,, \\
  u^t    &= [-g_{tt}-g_{\phi\phi}\Omega\,(\Omega-2\omega)]^{-1/2} \,.
\end{align*}
For an unperturbed Schwarzschild, the geodesic-orbit value of the latter is given by
\[(u^t_\pm)^{-2}=-g_{tt}-g_{\phi\phi}(\Omega_\pm)^2
                =\frac{(2r-M)^2-4Mr}{(2r+M)^2} \;.\]
Calculation of $u_{\phi,r}$ yields quite a cumbersome result, even after the restriction to linear perturbation, the more so after substituting the geodesic values $\Omega_\pm$, so we only illustrate it numerically below.

\subsection{Marginally bound circular geodesics}

It is also useful to check down to where the circular orbits (geodesics in our case) are energetically bound, i.e. having $-u_t\!<\!1$. In terms of the contravariant four-velocity components, the marginal case thus reads
\begin{align}
  1 &= u^t\left[e^{2\nu}+B^2\rho^2 e^{-2\nu}\omega(\Omega-\omega)\right] \nonumber \\
    &\doteq u^t\left(e^{2\nu_0}e^{2\nu_1}+B^2 r^2 e^{-2\nu_0}\omega_1\Omega\right),
\end{align}
where the second row restricts to the linear perturbation. In the second term, one can use $B^2 r^2 e^{-2\nu_0}\!=\!(2r+M)^4/(16r^2)$ and substitute just the unperturbed, pure-Schwarzschild value for the geodesic values of $\Omega$, thus finding
\[B^2 r^2 e^{-2\nu_0}\omega_1\Omega_{\pm}\doteq\pm\frac{1}{2}\,\sqrt{\frac{M}{r}}\,(2r+M)\,\omega_1 \,.\]

\subsection{Photon geodesics}
\label{photon-orbits}

The ``innermost'' possible tracks for free circular motion are found by equating the light-like limits of stationary circular motion (\ref{Omega,min,max}) with the geodesic values (\ref{Omega_pm})--(\ref{Omega_pm,approx2}). One thus obtains the conditions
\begin{align}
  (2r+M)^4\,\frac{\omega_{1,r}}{e^{2\nu_1}}
  = & \mp 8(2r-M)(2r-M-2\sqrt{Mr}) \nonumber \\
    & \pm 2\sqrt{\frac{r}{M}}\,(2r+M)(4r^2+M^2)\,\nu_{1,r} \,.
\end{align}
For a pure-Schwarzschild limit, they reduce to
\[(2r-M)^2=4Mr\]
which gives the correct Schwarzschild value.

\subsection{Coordinate versus geometrical measures}

The statements about space-times are often being expressed in coordinate terms, and this is especially the case when speaking of location of the important orbits we have just focused on. Although the isotropic, Weyl (or Schwarzschild) coordinates do represent some of the space-time features quite adequately, such statements should be made with caution. Specifically in the case of orbital radii, one should check them by employing more invariant measures, like proper radial distance or circumferential radius (given as proper circumference of a given circle divided by $2\pi$). This is mainly desirable in space-times where there is another source present like in situation we consider here, because this additional source also contributes to the metric and thus to the way how such measures are computed. Let us briefly check how it goes for our black-hole--disc system.

The proper radial distance from a horizon ($r\!=\!M/2$) to a given $r\!>\!M/2$, calculated along all the three remaining coordinates ($t$, $\theta$, $\phi$) constant, is given by
\[\int\limits_{M/2}^r\sqrt{g_{rr}}\,{\rm d}r
  =\int\limits_{M/2}^r e^{\zeta-\nu}\,{\rm d}r\,,\]
while the proper circumference corresponding to a certain $r$ (computed along constant $t$, $r$ and, in the equatorial plane, $\theta\!=\!\pi/2$) reads
\[\int\limits_0^{2\pi}\sqrt{g_{\phi\phi}}\,{\rm d}\phi
  =2\pi\sqrt{g_{\phi\phi}}
  =2\pi Bre^{-\nu}.\]

The proper distance would thus require to know $\zeta$ which is however quite a difficult task. Actually, this function is given, in the first perturbation order, by equations
\begin{align*}
  & (2\!-\!B)r\zeta_{,r}-B\zeta_{,\theta}\cot\theta+2\!-\!2B
       = B\left[r^2(\nu_{,r})^2\!-\!(\nu_{,\theta})^2\right], \\
  & (2\!-\!B)\zeta_{,\theta}+Br\zeta_{,r}\cot\theta-(2\!-\!2B)\cot\theta
       = 2Br\,\nu_{,r}\nu_{,\theta}
\end{align*}
(in the first order, the function $\omega$ does not enter at all), which could be only solved numerically (with our given $\nu$) and we have not done it in the first paper nor here.
On the other hand, the circumferential radius of the $r\!=\!{\rm const}$ rings is much easier to find,
\begin{equation}  \label{rcf}
  r_{\rm cf}=Bre^{-\nu}=\frac{(2r+M)^2}{4r\,e^{\nu_1}} \,,
\end{equation}
where, for the equatorial case, one substitutes the equatorial form of the disc potential for $\nu_1$.
Results obtained for the important circular orbits are shown in Figure \ref{orbits}, together with their coordinate values.

\subsection{Illustrations}

\begin{figure*}[ht]
\begin{center}
\includegraphics[width=\textwidth]{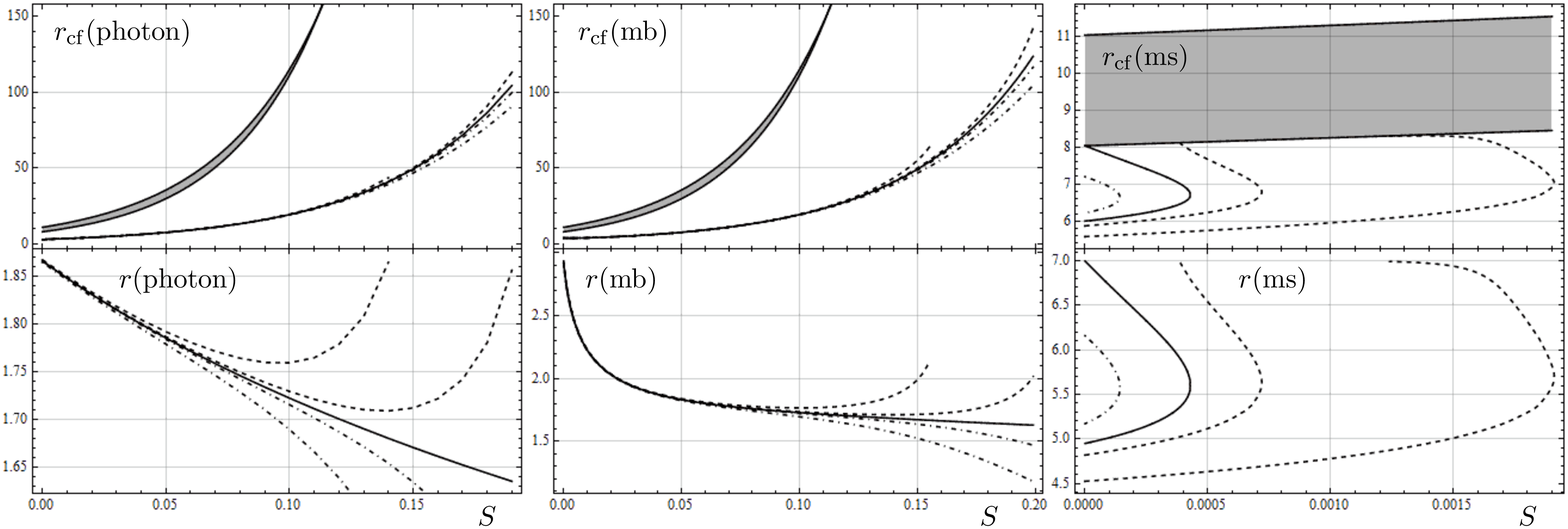}
\caption
{Dependence of the location of the photon (left), marginally bound (middle) and marginally stable (right) circular orbits on the disc mass (actually its density $S$), drawn for a disc existing between $r_{\rm in}\!=\!7M$ and $r_{\rm out}\!=\!10M$, for several values of the ``dragging density'' $W$. The bottom halves of the plots are drawn in terms of the isotropic radius $r$, while the top halves are given in terms of the corresponding circumferential radius $r_{\rm cf}\!\equiv\!\sqrt{g_{\phi\phi}}$. In all the plots, the middle (solid) line corresponds to $W\!=\!0$ and the two side curves correspond to $W\!=\!1/M$ and $W\!=\!5/M$; the orbits co-rotating with the disc are represented by dashed lines lying above the middle $W\!=\!0$ curve (or, to the right of it in the right plot), while the orbits counter-rotating with respect to the disc are represented by dot-dashed lines lying below the middle curve (to the left of it in the right plot; just one of these exists there). In the top plots, the region filled with the disc is shaded in grey. The radii are given in the units of $M$, the density $S$ is in the units of $1/M$.}
\label{orbits}
\end{center}
\end{figure*}

\begin{figure*}[ht]
\begin{center}
\includegraphics[width=\textwidth]{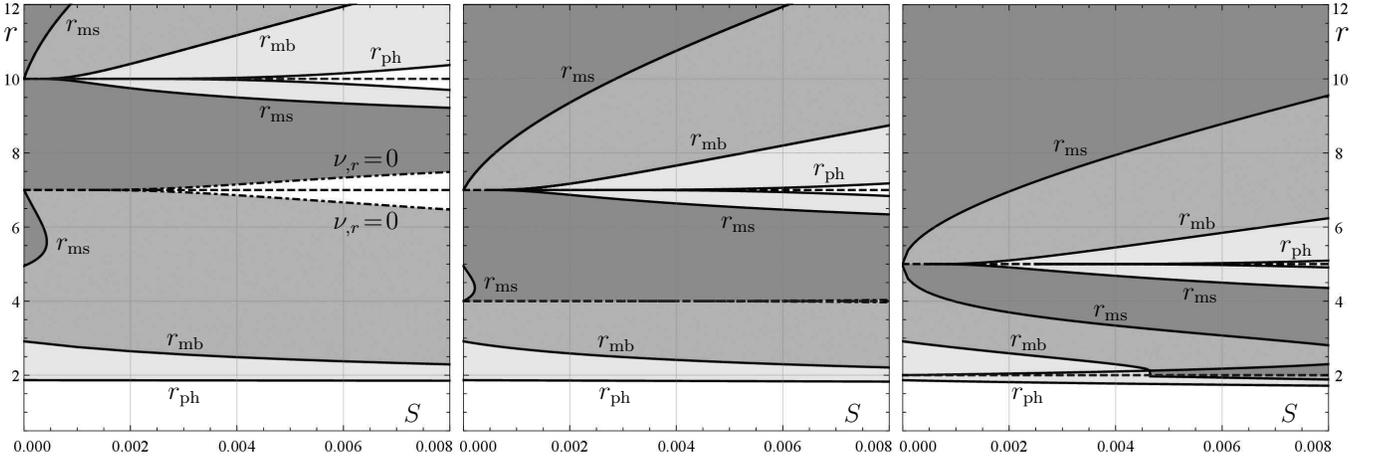}
\caption
{Dependence on the disc density $S$ of the isotropic radii $r$ of the photon ($r_{\rm ph}$), marginally bound ($r_{\rm mb}$) and marginally stable ($r_{\rm ms}$) circular geodesics, drawn, together with the (dot-dashed) boundaries of the region(s) where such orbits are possible as geodesics ($\nu_{,r}\!=\!0$), for the $W\!=\!0$ disc lying between $r_{\rm in}\!=\!7M$ and $r_{\rm out}\!=\!10M$ (left plot), between $r_{\rm in}\!=\!4M$ and $r_{\rm out}\!=\!7M$ (middle plot), and between $r_{\rm in}\!=\!2M$ and $r_{\rm out}\!=\!5M$ (right plot); the inner and outer radii of the discs are indicated by dashed lines. Meaning of the lines is clear from shading: dark shaded are regions where the geodesics are possible, time-like, bound and stable (in the horizontal direction), lighter grey are possible, time-like and bound but not stable, still lighter grey shows where they are only possible and time-like (but neither stable nor bound), and pure white indicates where they are only possible or where even none of the conditions is satisfied. It is helpful to keep in mind that for $S\!=\!0$, i.e. when there is no disc, all the graphs reduce to the pure-Schwarzschild values $r_{\rm ms}\!\doteq\!4.95M$, $r_{\rm mb}\!\doteq\!2.91M$ and $r_{\rm ph}\!\doteq\!1.87M$ (remember $r$ is the {\em isotropic} radius, not the Schwarzschild one). It is also seen that with increasing mass density the counter-rotating dust interpretation of the disc becomes problematic, because -- even for the disc on the largest radius (left plot) -- free circular orbits within the disc gradually cease to be stable/bound/possible/time-like. The radius is in the units of $M$ and the density is in the units of $1/M$.}
\label{orbits-shaded}
\end{center}
\end{figure*}

\begin{figure*}[ht]
\begin{center}
\includegraphics[width=\textwidth]{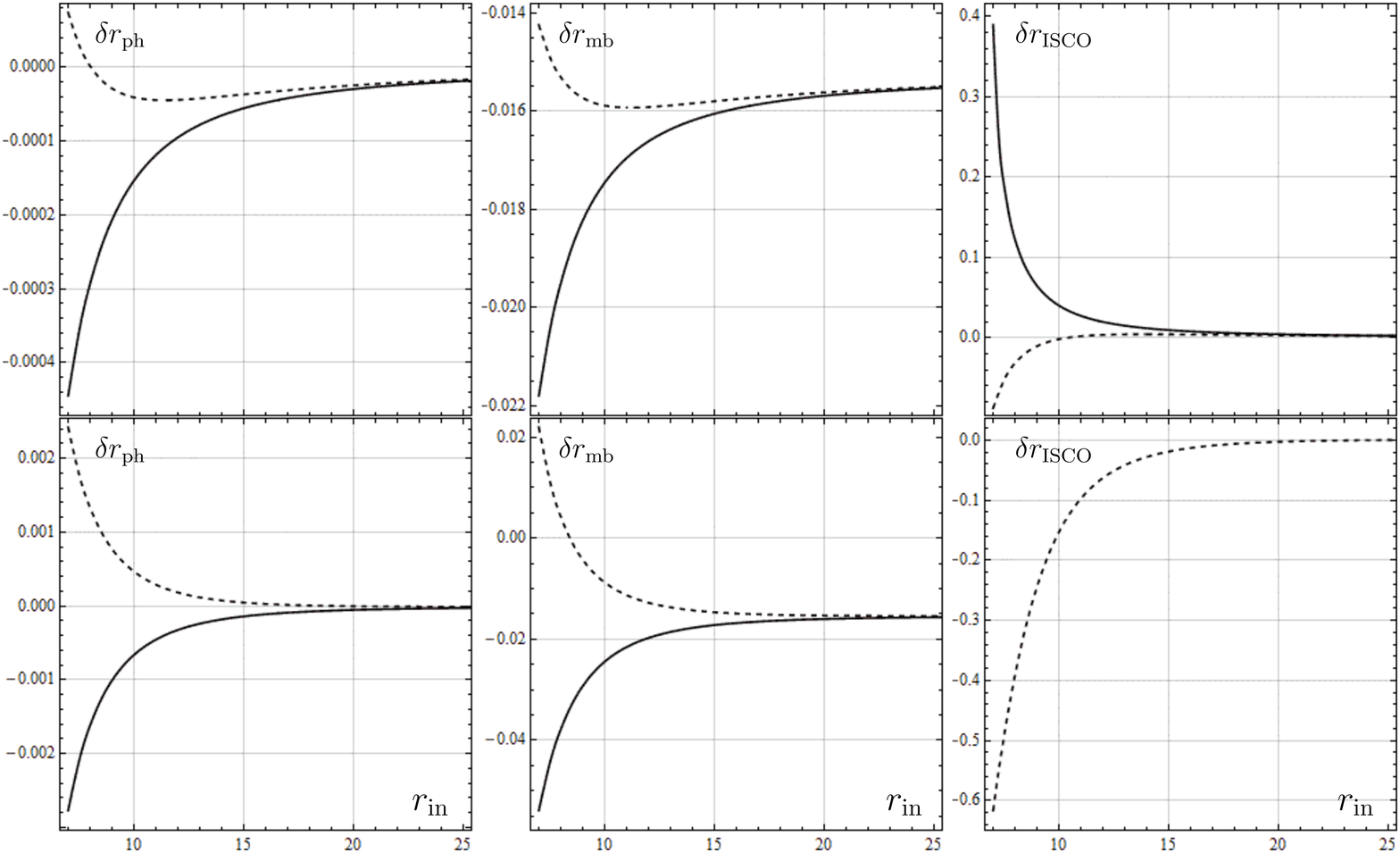}
\caption
{Dependence on the disc inner radius $r_{\rm in}$ of the isotropic radii of those photon ($r_{\rm ph}$), marginally bound ($r_{\rm mb}$) and marginally stable ($r_{\rm ms}\!=\!r_{\rm ISCO}$) circular geodesics which lie between the black hole and the disc, drawn for light discs (density $S\!=\!10^{-4}/M$) of radial width $r_{\rm out}\!-\!r_{\rm in}=3M$ and dragging densities $W\!=\!1/M$ (upper row) and $W\!=\!10/M$ (lower row). Actually plotted are differences between the radii of the respective co-rotating and counter-rotating orbits (drawn in solid/dashed lines) and the corresponding Schwarzschild values (marked as zeros on the axes) -- hence the notation by $\delta$. With increasing radius of the ring $r_{\rm in}$, the photon and marginally stable orbits approach their Schwarzschild positions, whereas the marginally bound orbit remains shifted due to the presence of the disc (orbital energy with respect to infinity is naturally affected by the disc for {\em any} $r_{\rm in}$; in addition, increasing the disc radius while keeping its density generally means increasing the disc mass, so its gravitational effect does not fall off as quickly as one might expect). All axes are scaled by $M$.}
\label{orbits-r}
\end{center}
\end{figure*}

In Figure \ref{orbits}, we exemplify the conditions derived above for a disc lying between $r_{\rm in}\!=\!7M$ and $r_{\rm out}\!=\!10M$, evaluating them numerically in terms of the coordinate (isotropic) as well as circumferential radii. The figure shows how the locations of the important circular equatorial geodesics depend on the disc mass (actually on its Newtonian density $S$) for several values of $W$. With increasing $S$, the photon and marginally bound orbits go down (from their Schwarzschild values) in coordinate radius, but their circumferential radii increase, because the exponential in (\ref{rcf}) ``beats'' the decrease of $r$. The interval of stable circular motion, existing between the disc and the pure-Schwarzschild location (in isotropic radius, it is $r\!\doteq\!4.95M$), shrinks and finally disappears for a certain $S$; for example, in the $W\!=\!0$ case this happens just above $0.0004/M$ (which corresponds to the total disc mass ${\cal M}_1\!\doteq\!0.064M$). Note that with increasing $S$ the circumferential radii of the disc inner and outer edges (the disc between them is grey shaded) first slightly recede from each other, but than come closer and finally ``intersect'' for $S\!\simeq\!0.11/M$. Well understandable from expression (\ref{rcf}), it indicates rather strong spatial curvature due to the potential valley generated by the disc. (Keep in mind that $S\!\simeq\!0.11/M$ is far beyond validity of the linear approximation -- see Section \ref{validity}.)

In Figure \ref{orbits-shaded}, the properties of free circular motion (both below, within and above the disc) are shown for discs lying at three different radii relatively close to the black hole. They are indicated by shades of grey: dark grey is where the geodesics are possible, time-like, bound and stable, lighter grey are possible, time-like and bound but not stable, still lighter grey shows where they are only possible and time-like (but neither stable nor bound), and pure white indicates where they are only possible or where even none of the conditions is satisfied. We remind the reader that ``possible'' means that there exists a value of angular velocity for which a circular track at a given radius is realizable as a geodesic. The boundary of a region where this is fulfilled (given by the Lagrangian circle $\nu_{,r}\!=\!0$) is indicated by the dot-dashed curve.

The plots shown in Figure \ref{orbits-shaded} are quite complicated, but one should bear in mind that i) their most complicated, right-hand parts are typically far beyond validity of the linear approximation (something like left quarter of the plots may be relevant, see Section \ref{validity}), and that ii) probably only the leftmost of the plots is reasonable astrophysically, because in the others the disc is too close to the black hole. Anyway, the best understandable and simply behaving are the orbits which exist {\em above} the disc: these represent photon, marginally bound and marginally stable orbits of the whole system; for low disc mass, the photon and the marginally bound orbits are seen to practically coincide with the edge of the disc. {\em Between} the black hole and the disc lie the orbits which ``belong to the black hole'' but shift from their Schwarzschild positions due to the disc presence. As expected, this shift is slow and gradual for the photon and marginally bound orbits, whereas the marginally stable orbit(s) are much more sensitive to the details of the field; focusing already on the disc entirely lying above the Schwarzschild radii of all the important orbits (leftmost plot), it is seen that there originally (for a negligible-mass disc) exists a region of stable circular geodesics between the Schwarzschild value of $r_{\rm ISCO}$ and the inner disc edge, but with increased disc mass this region shrinks and quite soon disappears, leaving the whole region below the disc unstable. In the middle plot, the inner edge of the disc lies {\em below} the pure-Schwarzschild radius of the ISCO, and the situation is seen to be just opposite to the previous case: there first exists an {\em unstable} region between the inner disc edge and the Schwarzschild value of $r_{\rm ISCO}$, which quite quickly shrinks and disappears with increased disc mass. In any case, the disc orbits quite soon (in the sense of increasing the disc mass) become unstable at the {\em outer} disc edge. And just a final note: interesting is the region around the mass value of $0.0005M$ in the middle plot, because there all the disc orbits are stable (and bound and time-like), although the inner disc edge is below the pure-Schwarzschild value of $r_{\rm ISCO}$.

The dependence of the important-orbit radii on the radius of the disc is plotted in Figure \ref{orbits-r}, specifically for the orbits lying between the horizon and the disc (the co-rotating as well as counter-rotating ones) and for small values of the disc mass density. With increasing disc radius, the orbits approach their Schwarzschild locations, except for the marginally bound orbits (determined by energy with respect to infinity and thus affected even when the disc lies on large radii).

\section{Physical requirements on the disc}
\label{phys-conditions}

In a previous paper \cite{CizekS-17}, we considered two interpretations of the disc -- the interpretation in terms of a single ideal fluid in stationary circular motion, characterized by surface density $\sigma$, azimuthal pressure $P$ and orbital velocity $v$, and the interpretation in terms of two counter-rotating but non-interacting (dust) streams orbiting the black hole on prograde and retrograde circular geodesics, characterized by surface densities $\sigma_\pm$ and orbital velocities $v_\pm$ (all the velocities are taken with respect to the local zero-angular-momentum observer, ZAMO). Let us supplement this part by listing basic physical requirements imposed on the disc and checking what they imply for the parameters of the above two pictures.

Let us remind, from the first paper, that the two interpretations correspond to writing the surface energy-momentum tensor
\[S^\alpha_\beta(\rho):=
  \int\limits_{z=0^-}^{z=0^+} T^\alpha_\beta e^{2\zeta-2\nu}{\rm d}z\]
in the forms
\begin{align*}
  S^{\alpha\beta}
  &=\sigma u^\alpha u^\beta+Pw^\alpha w^\beta \quad ({\rm 1~stream}) \\
  &=\sigma_+ u_+^\alpha u_+^\beta + \sigma_- u_-^\alpha u_-^\beta \quad ({\rm 2~streams}) \,.
\end{align*}
Here $u^\alpha$ is the ``bulk'' four-velocity,
\begin{align*}
  u^\alpha &= u^t (1,0,0,\Omega) \,, \\
  u_\alpha &= \rho B u^t \left(-\frac{e^{2\nu}}{\rho B}-\omega v,0,0,v\right) \,,
\end{align*}
where
\[(u^t)^2 =\frac{e^{-2\nu}}{1-B^2\rho^2 e^{-4\nu}(\Omega-\omega)^2}
          =\frac{e^{-2\nu}}{1-v^2}\]
and
\[v:=\rho B e^{-2\nu}(\Omega-\omega)=\sqrt{g_{\phi\phi}}\,e^{-\nu}(\Omega-\omega)\]
represents linear velocity with respect to the local ZAMO,
and $w^\alpha$ is the ``azimuthal'' vector perpendicular to $u^\alpha$, with components
\begin{align*}
  w^\alpha &= u^t\left(v,0,0,\frac{e^{2\nu}}{\rho B}+\omega v\right), \\
  w_\alpha &= \rho B u^t\,(-\Omega,0,0,1).
\end{align*}
The four-velocities $u_\pm^\alpha$ of the counter-rotating picture are of the $u_\pm^\alpha\!=\!u_\pm^\alpha (1,0,0,\Omega_\pm)$ form again, now with $\Omega_\pm$ (and the corresponding $v_\pm$) given by free circular motion (see Section \ref{free-circular}). The parameters of the two interpretations are related in a number of ways, which we mentioned in the previous paper.

\subsection{Energy conditions}

In terms of the surface energy-momentum tensor $S_{\mu\nu}$, the energetic conditions (slightly different requirements for the attractive character of gravity) read
\begin{align*}
  &{\rm weak~condition:} \quad S_{\mu\nu}\hat{u}^\mu\hat{u}^\nu \geq 0 \,,\\
  &{\rm dominant~condition:} \quad g_{\alpha\beta}{S^\alpha}_\mu{S^\beta}_\nu\hat{u}^\mu\hat{u}^\nu \leq 0 \,,\\
  &{\rm strong~condition:} \quad S_{\mu\nu}\hat{u}^\mu\hat{u}^\nu+\frac{S}{2} \geq 0 \,,
\end{align*}
where $\hat{u}^\mu$ represents any future-pointing time-like four-velocity ($\equiv$ a physical observer).
Substituting the one-stream expression for $S_{\mu\nu}$, one finds easily
\begin{align*}
  & S_{\mu\nu}\hat{u}^\mu\hat{u}^\nu = \hat{\gamma}^2(\sigma+\hat{v}^2 P) \,,\\
  & g_{\alpha\beta}{S^\alpha}_\mu{S^\beta}_\nu\hat{u}^\mu\hat{u}^\nu = \hat{\gamma}^2(-\sigma^2+\hat{v}^2 P^2) \,,\\
  & S_{\mu\nu}\hat{u}^\mu\hat{u}^\nu+\frac{S}{2} = \frac{\hat{\gamma}^2}{2}(\sigma+P)(1+\hat{v}^2) \,,
\end{align*}
where
\[\hat{\gamma}:=-u_\nu\hat{u}^\nu=\frac{1}{\sqrt{1-\hat{v}^2}}\]
is the Lorentz factor corresponding to the relative speed $\hat{v}$ of the fluid ($u^\mu$) with respect to the observer ($\hat{u}^\mu$).
Combining the two extreme cases $\hat{v}\!=\!0$ and $\hat{v}^2\!=\!1$ (plus, for the dominant condition, the requirement that the energy flow $-{S^\alpha}_\mu\hat{u}^\mu$ should be oriented towards future), we have, from the respective energy conditions,
\begin{align*}
  &{\rm weak:} \quad (\sigma\geq 0)\,\wedge\,(\sigma+P\geq 0) \,,\\
  &{\rm dominant:} \quad \sigma\geq |P| \,,\\
  &{\rm strong:} \quad \sigma+P\geq 0 \,.
\end{align*}
(Hence, the dominant condition implies the weak one, which even holds in general.)
For the two-stream interpretation, the energy conditions imply that the densities $\sigma_\pm$ be non-negative.

Physical intuition tells that the counter-rotating interpretation is only possible for discs with {\em non-negative} azimuthal pressure, $P\!\geq\!0$. Mathematically, this is seen from the trace of the energy-momentum tensor, $\sigma_++\sigma_-=-\sigma+P$. Together with the $P\!\geq\!0$ assumption, the requirements of the energy conditions can be summarized as
\begin{equation}  \label{energy-conditions}
  \sigma\geq P\geq 0, \quad \sigma_\pm\geq 0.
\end{equation}

Finally, let us recall how $\sigma_\pm$, $\sigma$ and $P$ came out for our example of constant-density disc, as given in equations (100) and (102), (103) in \cite{CizekS-17}: 
\begin{equation}
  \sigma_{\pm}=\frac{4r^2-8Mr+M^2}{8r^2}
               \left[S \pm \frac{W\,(Mr)^{3/2}}{2\pi\,(2r+M)^3}\right],
\end{equation}
\begin{align}
  \sigma & = +\Sigma +
              \sqrt{\Sigma^2 + \frac{16Mr\,(2r\!-\!M)^2\,\sigma_+\sigma_-}{(4r^2\!-\!8Mr\!+\!M^2)^2}} \;, \\
  P      & = -\Sigma +
              \sqrt{\Sigma^2 + \frac{16Mr\,(2r\!-\!M)^2\,\sigma_+\sigma_-}{(4r^2\!-\!8Mr\!+\!M^2)^2}} \;,
\end{align}
where we have denoted $\Sigma:=(\sigma_+\!+\!\sigma_-)/2$,
$S$ is the (constant) Newtonian surface density and $W$ is its counter-part (also constant) which appeared in integration of the dragging differential equation, both assumed to be positive and having the dimension of $1/M$.
Now the energy conditions (\ref{energy-conditions}) can be checked easily. Most notably, $\sigma\geq P\geq 0$ is clearly satisfied if $\sigma_\pm\geq 0$. The value of $\sigma_+$ only comes out negative at very low radii, namely at $r<(1+\sqrt{3}/2)M\doteq 1.866M$ (this is still above horizon which lies on $r=M/2$); the value of $\sigma_+$ might also turn negative elsewhere if $W$ were too large relative to $S$.

\subsection{Subluminal motion of the disc fluid}

Another obvious requirement is that the disc fluid should be moving with subluminal speed, $v^2\!<\!1$. In \cite{CizekS-17}, we got, for our constant-density disc,
\begin{equation}  \label{v^2}
  v^2 = \frac{4Mr\sigma-(2r-M)^2 P}{(2r-M)^2\sigma-4MrP} \;,
\end{equation}
where, when taking the square root (after substituting for $\sigma$ and $P$ from above), the $+/-$ sign should be chosen in case that $\sigma_+\!>\!\sigma_-\,$/$\,\sigma_+\!<\!\sigma_-\,$.
Combining the above requirement with $v^2\!\geq\!0$ and assuming that the disc lies above the photon circular geodesic of the original (Schwarzschild) space-time (namely where $(2r-M)^2>4Mr$), one obtains either $\sigma\!>\!|P|$, or $\sigma\!<\!-|P|$.
If considering the interpretation in terms of two counter-rotating geodesic streams, the whole disc should lie above both corresponding photon geodesics (Section \ref{photon-orbits}).

Note that in {\em astro}physical settings the requirement of subluminal motion is not sufficient, namely, if adhering to the two-stream interpretation, one would also add that the whole disc should lie where the free circular motion is stable (Section \ref{ms}).

\section{Concluding remarks}

Several important properties have been derived of space-times generated by a black hole surrounded, in a symmetric manner, by a rotating light finite thin disc. We have thus continued our study of the corresponding linear perturbation of Schwarzschild solution presented recently in \cite{CizekS-17}. Due to the disc, the black-hole horizon grows bigger and oblate, inflating towards the external source as usual. No ergosphere occurs in the first perturbation order and the central singularity remains like in the original, Schwarzschild space-time. Free circular equatorial motion {\em is} affected by the presence of the disc, as best illustrated by how the radii of important orbits change with the disc mass and radius. Since these orbits (mainly the innermost stable one, ISCO) are crucial for disc-accretion scenarios, their shift due to the disc's own gravity should indicate how a real accretion configuration might differ from its test-matter approximation. For some parameter ranges this implies just shift of the inner edge of the accretion disc, while for other it might indicate a tendency for radial fragmentation. Finally, to calculation of basic physical parameters of the disc (made in previous paper), we have added a check of the natural physical requirements like energy conditions or time-like (subluminal) character of disc-matter motion; they lead to simple conditions for parameters in terms of which the disc is interpreted.

At several places, we compared the present results/figures with those obtained e.g. in \cite{Erratum-01,Semerak-03}. The whole series to which these older papers belong (as well as others' work cited therein) studied the influence of disc or ring matter configurations on space-time of a black hole within {\em static} and axisymmetric class of metrics. In the static case, the problem is much easier than in the more general, stationary case considered here, because then $\omega\!=\!0$ and the potential $\nu$ superposes linearly. In the present paper, stationary rotation of the sources {\em is} admitted, though it is only taken into account in the linear perturbation order. We have seen that in this order the properties which are given solely by the gravitational potential ($\nu$) keep their static (Schwarzschild-like) form, because rotation (dragging angular velocity $\omega$) only ``back-affects'' the potential in the second order.

The main literature on gravitating stationary (rotating) discs or toroids around black holes was already mentioned in the first paper. In particular, we summarized the paper by \cite{Lanza-92} who considered a very similar task (a thin disc around a black hole, assuming stationarity and axial symmetry), but solved it ``exactly'' (numerically), with a different kind of disc and under different assumptions. We thus concluded there that it is difficult to compare Lanza's results directly with our linear-perturbation approximation, although in the limit of very light disc the results should be similar in some sense; we plan to return to this point in future. However, to mention at least one clear point, in contrast to the above paper, we have never obtained a prolate deformation of the black-hole horizon.
Let us, in addition, refer here to the paper by \cite{BarausseRPA-07} which treated, numerically, a {\em thick} toroid rather than a thin disc (around a rotating black hole), but whose point is very close to that of ours: to provide a space-time describing a {\em reasonable} deviation from Kerr and test its implications for astrophysical black-hole sources. In particular, they checked there whether the presence of the massive toroid somehow changes the gravitational waveforms produced by an equatorial inspiral of a small body onto such a system (and found that the effect in general is very small).

Finally, possible future plans include trying to compute the first-order perturbation due to a different type of disc. Actually, the disc with a constant density is probably not very realistic astrophysically, although one can hardly expect to be able to integrate the Green functions over a more generic function of radius, specifically for some which would really follow from some model of high-angular-momentum accretion.
Another challenge is of course to extend the treatment to higher perturbation orders. Namely, in the linear order there is no back reaction of the disc to its own gravity (no {\em self}-gravity), so the most inherent feature of general relativity is not present (specifically, the potential superposes like in the static case, because dragging only enters the equation for potential quadratically). Unfortunately, the differential equations for the $k$-th perturbation terms have $(k\!-\!1)$-th (and lower-order) terms on their right-hand sides, so it is practically impossible to find their solution if the right-hand sides are not very simple; the latter is clearly not our case.

\acknowledgments
We are grateful to N. G\"urlebeck and T. Ledvinka for interest and useful comments.
O.S. thanks for support from the grant GACR-17/13525S of the Czech Science Foundation.

\bibliography{perturbace-2.bib}

\end{document}